\def\be{\begin{equation}}
\def\ee{\end{equation}}
\def\bea{\begin{eqnarray}}
\def\eea{\end{eqnarray}}
\def\ba{\begin{array}}
\def\ea{\end{array}}
\def\nm{\nonumber \\ }

\def\b{\beta}
\def\bm{\beta ^{-1}} 
\def\b2{\beta ^2}

\def\vf{\varphi}
\def\k{\kappa}
\def\l{\lambda}
\def\L{\Lambda}

\def\a{\alpha}

\def\e{\epsilon}
\def\S{\Sigma}

\def\Z{Z \! \! \! Z}

\def\v{\vert 0 \rangle }

\def\is#1{\vert #1 \rangle}

\def\W{{\cal W}}
\def\wa2{$W\! \! A_2$}
\def\wbc2{$W\! BC_2$}
\def\wg{$W\!G \ $}

\def\d{\partial}
\def\sp{\qquad ; \quad }
\def\s{\ , \ }

\def\p{\partial}

\def\kh{(2\leftrightarrow 3)}

\def\lra{\leftrightarrow}

\def\NPB#1#2#3{{\rm Nucl. Phys.} {\bf B#1} (#2) #3}
\def\PLB#1#2#3{{\rm Phys. Lett.} {\bf #1B} (#2) #3}

\def\IJMPA#1#2#3{{\rm Int. J. Mod. Phys.} {\bf A#1} (#2) #3}
\def\IJMPB#1#2#3{{\rm Int. J. Mod. Phys.} {\bf B#1} (#2) #3}   

\def\CMP#1#2#3{{\rm Commun. Math. Phys.}{\bf #1} (#2) #3}
\def\PR#1#2#3{{\rm Physics Reports}{\bf #1} (#2) #3}
\def\SCRAP#1#2#3{{\rm Sov. Sci. Rev. A. Phys.}{Vol. #1} (#2) #3}
\def\TMP#1#2#3{{\rm Theor. Math. Phys.}{\bf #1} (#2)  #3}
\def\AP#1#2#3{{\rm Ann. of Phys.}{\bf #1}(#2) #3}

\def\JPG#1#2#3{{\rm J. of Geom. and Phys.}{Vol. #1},n.#2,#3}

\documentstyle[12pt]{article}

\advance \topmargin by -\headheight
\advance \topmargin by -\headsep

\evensidemargin \oddsidemargin
\begin{document}

\title{  \begin{flushright}
\normalsize{ ITP Budapest Report No. 524 }
  \end{flushright}
  \vspace{5cm}
  A new approach to the correlation functions of  $W$-algebras      }
       \author{ \large{ Z. Bajnok } \\ \\
       \normalsize {\it Institute for Theoretical Physics }\\
       \normalsize {\it  Roland E\"otv\"os   University, }\\
       \normalsize {\it H-1088 Budapest, Puskin u. 5-7, Hungary}}

       \def\today{ April 26, 1997 }

\maketitle

\vspace{1.5cm}

\begin{abstract}

We propose a new approach to the study of the correlation functions of  
$W$-algebras. The conformal blocks (chiral correlation functions), 
for fixed arguments, are defined 
to be those linear functionals on the product of the highest weight (h.w.)
representation spaces which satisfy the Ward identities. 
First we investigate the dimension of the chiral correlation functions in 
the case when there is no singular vector in any of the representations. 
Then we pass to the analysis of the completely degenerate representations.
A special subspace of the h.w. representation spaces, introduced by Nahm, 
plays an important role in the considerations. The structure of these 
subspaces shows a deep connection with the quantum and classical Toda models 
and relates certain completely degenerate representations of the \wg 
algebra to representations of $G$. This is confirmed by an
analysis for the Virasoro, \wa2 and 
\wbc2 algebra. 
We also relate our work to Nahms, Feigen-Fuchs' and Watts' results.  

\end{abstract}

\section{ Introduction}

Two dimensional conformal field theories have  wide range 
of applications. They describe the critical behaviour of 
two dimensional statistical physical models since at the 
second order phase transition point the theory is scale 
and consequently conformal invariant. The conformal invariance 
in string theory arises as a remnant of the reparametrization 
invariance of the worldsheet which guarantees that the physics 
does not depend on the coordinate choices. The success of 
conformal field theories and the reason of their mathematical 
applications are due to the fact that they possess infinite dimensional
symmetry algebras, which help to solve the models. These 
algebras necessarily contain the conformal or Virasoro algebra, 
in most cases however, they are larger. The extensions of the 
conformal algebra are called $W$-algebras.  

The analysis of the $W$-symmetric theories, theories whose 
symmetry algebra is a $W$-algebra,  can be divided into 
two steps. The first concerns the investigation of the symmetry
algebra itself. It contains the presentation of the algebra, 
which can be done by commutation relations or by localizing 
it as a subalgebra in a known algebra, etc.  . Then the representation
theory of the symmetry algebra has to be developed concentrating
on the irreducibility of the representations and the singular 
vectors. The next step is the solution of a model with a given
$W$-symmetry. This means the determination of the correlation functions, 
which is usually done either by solving the differential equations  that 
arises from the decoupling equations of the singular vectors, or 
by using some auxiliary field technics. 

In most cases the analysis of $W$-symmetric theories was focused on 
the symmetry algebra and their representation theory on one hand 
 \cite{BouShou}, 
or concentrated on auxiliary constructions which work well in principle 
only for rational theories, on the other hand \cite{BiW,FaLuW}. 
In the string theoretical applications, however, nonrational and rational
theories are equally important. In Calabi-Yau theories the theory and 
consequently the symmetry algebra depend on several parameters, which not
necessarily take rational values. 

In this paper we concentrate on the second step, ie. we try to analyse 
the correlation functions of general $W$-symmetric models. For this we give 
a new definition for the conformal blocks, placing emphasis on the
generality and calculability.   

The definition of the chiral correlation functions or conformal 
blocks presented here is the adaptation what was used by Felder for 
Kac-Moody algebras in \cite{FelLH}. In principle the conformal 
blocks are defined to be those linear functionals on the product of the
highest weight (h.w.) representation spaces that respect the Ward
identities. The simple idea behind this can be understood in the following way:
consider  a correlation function of some descendant field in 
the Virasoro theory 
\be
\langle \Phi_1(z_1)\dots {\cal L}_n\Phi_i(z_i) \dots
\Phi_N(z_N)\rangle =\oint_{C_i}{dz\over 2\pi i}
\langle \Phi_1(z_1)\dots (z-z_i)^{n+1}L(z)\Phi_i(z_i)
\dots \Phi_N(z_N)\rangle \s n<0  ,
\ee
where all insertion points, $\{ z_i \}$, are different
and none is zero or infinite.
Here and from now on $C_i$ denotes  a small integration contour 
around $z_i$. Now we can make a usual contour deformation. Observe that  
$n<0$ so there is no pole at infinity and at zero (this is also true for 
$n<2$ due to the quasi primary nature of $L(z)$). Consequently  
we obtain the following formula:
\be
\langle \Phi_1(z_1)\dots {\cal L}_n\Phi_i(z_i) \dots \Phi_N(z_N)
\rangle =-\sum_{j:j\neq i}\oint_{C_j}{dz\over 2\pi i}\langle 
\Phi_1(z_1)\dots (z-z_i)^{n+1}L(z)\Phi_j(z_j) \dots \Phi_N(z_N)\rangle .
\label{ward}
\ee
Expanding $(z-z_i)^{n+1}$ around $z_j$ we get an infinite sum, each term
of it being associated with an action of the Virasoro algebra. Since 
we are working with h.w. representations the sum is finite and we 
end up with a well-defined action. Denoting the sum by $\tilde {\cal L}_n$ 
we have:
\be
\langle \Phi_1(z_1)\dots {\cal L}_n\Phi_i(z_i) \dots \Phi_N(z_N)
\rangle =-\sum_{j:j\neq i}\langle \Phi_1(z_1)\dots 
\tilde {\cal L}_n\Phi_j(z_j) \dots \Phi_N(z_N)\rangle  .
\ee
In the singular terms of the operator product expansion we can replace
${\cal L} _{-1}\Phi_j(z_j) $  by $ \p _{z_j} \Phi_j(z_j)$ and
${\cal L} _0\Phi_j(z_j) $  by $ h_j \Phi_j(z_j)$ and recover the 
well-known Ward identities. (This relation will be used to define the 
conformal blocks later). 

Doing the same in the \wa2 theory we realize a problem  which concerns 
the modes $W_{-1}$ and $W_{-2}$: consider a correlation function of some 
$W$-descendants in a model with \wa2 symmetry:
\be 
\langle \Phi_1(z_1)\dots {\cal W}_n\Phi_i(z_i) \dots 
\Phi_N(z_N)\rangle =\oint_{C_i}{dz\over 2\pi i}
\langle \Phi_1(z_1)\dots (z-z_i)^{n+2}W(z)\Phi_i(z_i) 
\dots \Phi_N(z_N)\rangle \s n<0   .
\ee
Next we do the same contour deformation manipulations as before and expand the 
$(z-z_i)^{n+2}$ term around $z_j$. The sum gives rise to a well-defined 
operator on h.w. $W$-primary fields which is denoted by $\tilde{\cal W}_n$ 
with which the result is:
\be
\langle \Phi_1(z_1)\dots {\cal W}_n\Phi_i(z_i) \dots \Phi_N(z_N)
\rangle =-\sum_{j:j\neq i}\langle \Phi_1(z_1)\dots 
\tilde {\cal W}_n\Phi_j(z_j) \dots \Phi_N(z_N)\rangle  .
\label{wardw}
\ee
There is an important difference compared to the Virasoro theory. 
Namely in the singular terms  of the OPE one can only replace 
${\cal W} _0\Phi_j(z_j) $ by $w_j\Phi_j(z_j)$, but in principle one 
cannot relate the descendant fields ${\cal W} _{-1}\Phi_j(z_j) $  and 
${\cal W} _{-2}\Phi_j(z_j) $ to the primary field $\Phi_j(z_j)$. 
In the Virasoro theory $L_{-1}$ has a nice interpretation as the 
differentiation operator however this fails in the $W$-case. 

What to do now? 

We will give up in some sense the $L_{-1}=\p$ relation and reformulate
the technics described above in our new framework. We will see that relevant 
algebraic questions like the dimension of the conformal blocks, or the 
vanishing of the two and three point functions (the possible couplings),
can be analyzed without using the $L_{-1}=\p$ relation.

We start with the Virasoro symmetric models in Section 1. The aim of this
section is to reexplorer the well-known results of the conformal symmetric
models in order to be getting acquainted with the new framework. In doing so
we define the space of the conformal
blocks to be the space of those linear functionals on the 
product of the h.w. representation spaces which satisfy  the Ward identities
(\ref{ward}). Then the dimension of the two point functions is investigated, 
and it is shown that it is one if the conformal dimensions coincide 
and zero otherwise. In the case of the three point functions we analyse
various things. The dimension of the space of the three point functions
is one for arbitrary representations. If one of the representations has 
a singular vector than fixing one from the remaining weights and supposing
non vanishing three point functions we can determine the possible values for 
the last weight. We also comment on the fusion defined by Feigin and Fuchs.  
In the case of the four point function we show that its dimension is infinite
in general. Moreover if there is a singular vector in any of the representation
at level $n$ then the dimension of the space of the conformal blocks is not 
greater than $n$. After considering the general $n$-point functions we define
the $z$-dependence of the conformal blocks via the Friedan-Shenker connection. 
Geometrically the conformal blocks are horizontal sections of a holomorphic 
bundle. Although this approach naturally generalizes to higher genus 
surfaces we restrict our attention to the sphere. In this paper we focus on the 
algebraic and not on the geometric aspects and on the way it can be 
extended to general $W$-algebras. 

In generalising we start with the simplest $W$-algebra, namely with theories
having the \wa2 algebra, \cite{Zamalg}, as the symmetry algebra in Section
2.  We define the 
space of the conformal blocks in an analogous manner namely by means of 
the Ward identities. Then the dimension
of the space of the correlation functions is investigated. It is shown 
that a nonvanishing two point function implies that the Virasoro weights 
are the same and the $W$-weights are opposite. In the case of the 
three point functions even the dimension of the space of the conformal 
blocks is infinite in the general case. In analysing the three point 
functions it turns out that a "special" subspace of the h.w.
representation spaces plays a very important role. This space is a 
factor space  spanned
by those negative modes that annihilate the vacuum modulo those that 
do not.

In the case of a general $W$-algebra we give the definition of the 
conformal blocks and analyse the dimension of the space of the 
correlation functions. The $z$-dependence of the correlation 
function is defined similarly to the Virasoro case. 

In any of the consideration the special subspace of the h.w. representation 
space plays a crucial role. If one associate for a h.w. representation of 
$G$ a h.w. representation of \wg, then the dimension of the special 
subspace coincides with the dimension of the underlying representation.  
This connection can be made more explicit with the help of the singular 
vectors of the representation and by exploiting the relation with the 
classical Toda models. This is illustrated in section 4. 

The relation between the classical and quantum representations is supported 
by the example of the $B_2$ and $C_2$ Toda models which are given in the 
appendix.

\section{ The Virasoro theory}

Let $\S $ be the Riemann sphere with $N$ distinct points $p_1, 
\dots, p_N$. For the point $p_i$ we associate an irreducible 
h.w. representation of the Virasoro algebra, $V_i$, and denote 
the product of the representations by $V=V_1\otimes V_2\otimes \dots
\otimes V_n$. We choose coordinate functions
on the two covering neighbourhoods on the sphere: $z$ and $w=1/z$ and
write $z(p_i)=z_i$. Introduce the space of meromorphic vector fields
on the sphere with poles at $z_i$:
\be
\W^2(\S\setminus \{p_i\})=
\left \{ f(z)\ {\rm meromorphic} \ \bigl \vert \quad
f(z){d\over dz}\ {\rm  holomorphic \  except } \  z_1,\dots, z_N \right \}.
\ee
(We note that ${d\over dz},z{d\over dz},z^2{d\over dz}$ are holomorphic 
everywhere.) In the following we use the abbreviation 
$\W^2=\W^2(\S\setminus \{p_i\})$. For each function $f\in \W^2$ 
one can associate an action of the Virasoro algebra on $V$ in the following way. 
Make a Laurent expansion of $f(z)$ around $z_i$ and for the $(z-z_i)^{k+1}$ 
term associate the action of $L_k$ on  $V_i$ and consequently the action of 
$1\otimes \dots \otimes L_k \otimes \dots \otimes 1$ on $V$, for which we write
$L_k^{(i)}$. Since the functions of $\W^2$ are spanned by the functions of the 
form $(z-z_i)^{n} \ , n< 3$, we give the action explicitly for these:
\be
(z-z_i)^{-n}\rightarrow L^{(i)}_{-n-1}+ \sum_{l:l\neq
i}  \tilde L^{(l)}_{-n-1} ,
\ee
where we have to make a difference depending on whether $n>0$ or $n\leq 0$.
In these cases 
\be
\tilde L^{(l)}_{-n-1}=\left \{ \ba {ll} (-1)^n \sum_{k=0}^\infty
z_{il}^{-n-k}
{\scriptstyle{n+k-1\choose n-1}}L^{(l)}_{k-1} & n>0 \\
(-1)^n \sum_{k=0}^n z_{il}^{n-k} {\scriptstyle{n \choose
k}}L^{(l)}_{k-1} & n\leq 0  \ea \right .    
\ee
Although we have an infinite sum in $\tilde L^{(l)}_{n-1}$ only
finitely many terms contribute when it acts on h.w. representation
spaces. We note that any mode that appears in  $\tilde L^{(l)}_{n-1}$ 
annihilates the vacuum and its index is grater than $n-1$ (if  $n<-1$). 

The conformal blocks or chiral correlation functions assign a complex
number to fixed insertion points $\{z_i\}$ and associated representations
$V_i$, in such a way, that the Ward identity (\ref{ward}) holds. 
Reformulating this we define the space of the conformal blocks 
associated to $V$ and $\{p_i\}$ to be the space of those  
linear functionals on $V$ that is annihilated by $\W^2$ (for the 
natural right action), ie.
\be
E\left (V,\{p_i\}\right )=\left \{ u\in V^* \ \vert \quad u(xv)=0 \ ;
\quad \forall v\in V \ , \ x\in \W^2 \right \} .
\ee
We will vary the insertion points and consider the $z$-dependence later, 
now we focus on the dimension of the space of the conformal blocks. 

We start with the two point function $\langle \vf_1(z_1) \vf_2(z_2) \rangle $.
Consider a conformal block $u$ acting on a general element
\be
v=L_{-n_1}\dots
L_{-n_k}\is{\vf_1}\otimes  L_{-m_1}\dots L_{-m_l}\is{\vf_2}
, \ \  n_i\geq n_{i+1} , \  m_i\geq m_{i+1}
\ee
of the product of the h.w. representation spaces. 
(The h.w. representation spaces are graded, the eigenvalue of the 
operator $L_0$ is the level. 
This grading naturally extends to $V$: the levels sum up). 
Now if $v$ contains a mode 
$L_{-n}\s n>1$ then without loss of generality we have
\be
v=\left (L^{(1)}_{-n_1}+\tilde L^{(2)}_{-n_1}\right )v^{'}-
\tilde L^{(2)}_{-n_1}v^{'},
\ee
where clearly $ v^{'}=L_{-n_2}\dots
L_{-n_k}\is{\vf_1}\otimes  L_{-m_1}\dots L_{-m_l}\is{\vf_2}$.
Since $u$ annihilates the first term $u(v)=u( v^{''}) $, where $v^{''}=
-\tilde L^{(2)}_{-n_1}v^{'}$.  Note that the terms in  $v^{''}$ 
have level smaller than the level of $v$.  This means that the value of 
the conformal block acting on a general element can be expressed with its
values acting on vectors with smaller level. Since we are working with 
h.w. representation spaces the levels are bounded from below, so applying 
this procedure from 
level to level we can eliminate all the modes, which are not $L_{-1}$.
Consequently we analyse the case of the vector $v$ of the form  
$v=\left (L_{-1}\right )^n\is{\vf _1} \otimes
\left (L_{-1}\right )^m\is{\vf _2}$. 
Up to now we have not used the constraints coming from the everywhere
homomorphic (global) transformations, which are:
\bea
u\left ((L^{(1)}_{-1}+\tilde
L^{(2)}_{-1})v \right )&=&u\left ((L^{(1)}_{-1}+L^{(2)}_{-1})v \right ) =0\nm
 u\left ((L^{(1)}_{0}+\tilde
L^{(2)}_{0})v \right )&=&u\left ((L^{(1)}_{0}+L^{(2)}_{0}+z_{21}L^{(2)}_{-1})v 
\right )
=0\\  u\left ((L^{(1)}_{1}+\tilde
L^{(2)}_{1})v \right )&=&u\left ((L^{(1)}_{1}+L^{(2)}_{1}+2z_{21}
L^{(2)}_{0}+z_{21}^2L^{(2)}_{-1})v \right ) =0 \nonumber .
\label{vir2}
\eea
Two of the equations above can be used to eliminate all the $L_{-1}$-s. Using
an appropriate combination of all the three we have one more constraint, 
which reads on $v=\is{\vf_1}\otimes \is{\vf_2} $ as:
\be
u\left (\left (L^{(1)}_{1}+L^{(2)}_{1}+z_{21}(-L^{(1)}_{0}+L^{(2)}_{0}
)\right )v \right )=
 z_{21}(L^{(2)}_0-L^{(1)}_0)u\left (v \right )=0  .
\label{virh=}
\ee
This means however, that fixing one of the $L_0$ eigenvalues, say 
$L^{(1)}_{0}\is{\vf_1}=h_1$, we have only the $L^{(2)}_{0}\is{\vf_2}=h_2=h_1$
possibility for the other. Consequently the dimension of 
the space of the chiral two point function is one if the conformal 
weights coincide and zero otherwise. 

Now we analyse the three point functions with insertion points  $z_1,z_2,z_3$.
The argument used above to eliminate the modes $L_n \s n<-1$ generalises 
naturally for this case, so it is enough to take into account the elements of the h.w. 
representation spaces generated by acting $L^{(i)}_{-1}$ on the h.w. 
vectors. The global transformations restrict this space as:
\bea
u\left ((L^{(1)}_{-1}+\tilde L^{(2)}_{-1}+\tilde L^{(3)}_{-1})v \right )&=&
u\left ((L^{(1)}_{-1}+L^{(2)}_{-1}+\kh) v \right ) =0\nm
u\left ((L^{(1)}_{0}+\tilde L^{(2)}_{0}+\tilde L^{(3)}_{0})v \right )&=&
u\left ((L^{(1)}_{0}+L^{(2)}_{0}+z_{21}L^{(2)}_{-1}+\kh) v \right ) =0\\
u\left ((L^{(1)}_{1}+\tilde L^{(2)}_{1}+\tilde L^{(3)}_{1})v \right )&=&
u\left ((L^{(1)}_{1}+L^{(2)}_{1}+2z_{21}
L^{(2)}_{0}+z_{21}^2L^{(2)}_{-1}+\kh) v \right ) =0 \nonumber  .
\label{vir3}
\eea
Now we can make different interpretations. We can fix say all the three 
$L_0$ eigenvalues. In this case all the $L^{(i)}_{-1}$-s can be 
eliminated, and no other constraints remain. This shows 
that the dimension of the chiral three point function is one for 
all possible h.w. vectors. 

In order to interpret the results in another way we can express 
$L_0^{(1)}$ as:
\be
z_{21}L_0^{(1)}=L_1^{(1)}+L_1^{(2)}+L_1^{(3)}+z_{21}L_0^{(2)}+(z_{31}+z_{32})
L_0^{(3)}+z_{31}z_{32}L_{-1}^{3}
\label{L0}
\ee
Now we can fix the eigenvalue $L_0^{(2)}=h_2$ and $L_0^{(3)}=h_3$ and 
investigate the possible weights $L_0^{(1)}$ that can couple two them. We 
see that there are infinitely many possible values for $L_0^{(1)}$, 
the symmetry 
does not restrict the possible couplings.  Of course this will change 
drastically if we have singular vectors in the third representation space. 
We come back later to investigate this. 

Now consider the case of the $N$ point functions with insertion 
points: $z_1, z_2, \dots z_N$. The analysis follow the same line
we used for the two and three point functions. We show that the
value of the conformal block, $u$, on the general vector 
\be
v=L_{-n_1}\dots L_{-n_k}\is{\vf_1}\otimes \dots \otimes
L_{-m_1}\dots L_{-m_l}\is{\vf_N},
\ee
(where $ n_i\geq n_{i+1} \s m_i\geq m_{i+1}$),
can be re-expressed in terms of its values on vectors at lower level, 
if $v$ contains operators $L_{-n}, n>1$. Suppose that at the position $i$ 
we have a mode $L_{-j-1}\s j>0$, ie.  $v=L^{(i)}_{-j-1}v^{'}$. In this case, 
using the defining relations of the conformal blocks, we get 
$u (v)=u (v^{''})$ where $v^{''}=-\sum_{l:l\neq i}\tilde
L^{(l)}_{-j-1}v^{'}$. Each term in $v^{''}$ has smaller level than the level of $v$. 
Following this procedure we can eliminate all the modes inductively in the
level, which are not $L_{-1}$. The induction becomes complete since 
at the lowest levels we have vectors containing only $L_{-1}$-s. 
This shows that it is enough to define the value of the conformal blocks 
on the vectors containing only $L_{-1}$. They are not all independent
however, since we have the global Ward identities:
\bea
u\Bigl ((L^{(i)}_{-1}+\sum_{l:l\neq i}\tilde L^{(l)}_{-1})v \Bigr )&=&
u\Bigl ((L^{(i)}_{-1}+\sum_{l:l\neq i}L^{(l)}_{-1}) v \Bigr ) =0\nm
u\Bigl ((L^{(i)}_{0}+\sum_{l:l\neq i}\tilde L^{(l)}_{0})v \Bigr )&=&
u\Bigl ((L^{(i)}_{0}+\sum_{l:l\neq i}(L^{(l)}_{0}+z_{li}L^{(l)}_{-1}))
v \Bigr ) =0\\   
u\Bigl ((L^{(i)}_{1}+\sum_{l:l\neq i}\tilde L^{(l)}_{1})v \Bigr )&=&
u\Bigl ((L^{(i)}_{1}+\sum_{l:l\neq i}(L^{(l)}_{1}+2z_{li}
L^{(l)}_{0}+z_{li}^2L^{(l)}_{-1})) v \Bigr ) =0 \nonumber    .
\label{virN}
\eea
By the help of these equations we can eliminate the $L_{-1}$ modes in 
three arbitrary representations, however the others remain undetermined. 
This shows that the dimension of the space of the $N$-point functions is
infinite in the general case. (Sometimes we omit to write out the 
conformal block, $u$, in relations like above).  

Focusing on the four point case we have to define 
the values of the conformal blocks on the vectors:
$\is{\vf_1}\otimes \is{\vf_2}\otimes \is{\vf_3}\otimes
\left (L_{-1}\right )^n\is{\vf_4}$. 
This space is infinite dimensional in the general case. If however the
representation corresponding to $\vf_4$ is degenerate, then it contains
a singular vector at some level, say $n$ , of the form   
$\left ( \left (L_{-1}\right )^n+\dots \right )\is{\vf_4}=0 $. 
This shows that it is enough to 
define $u_i=u\left ((L_{-1}^{(4)})^{i}v \right ) \s i<n$, all others can be
re-expressed in terms of these, ie. the dimension of 
the space of the conformal blocks is $n$ 
in this case, which is the level of the singular vector. 

Now consider the effect of the singular vectors in the case of  the three point
functions. In equation (\ref{L0}) $L_0$ becomes an $n\times n$
 matrix since $\left ( L_{-1}^{(3)}\right ) ^n$ can be expressed in
terms of $\left ( L_{-1}^{(3)}\right ) ^k \s k<n$. Fixing 
$L_0^{(2)}=h_2$ we can diagonalize $L_0^{(1)}$ to obtain the possible 
nonvanishing couplings. Clearly the number of the nonzero couplings is
bounded by the level of the singular vector. 

We analyse the degenerate representaions starting from the simplest cases. 
The  simplest degenerate representation is the vacuum representation
it contains a singular vector at level one of the form $L_{-1}\is{\vf_3}=0$.
Since $L_i^{(3)}\is{\vf_3}=0$ for $i=-1,0,1$ then the analysis of the
possible couplings reduces for the same analysis, what we have performed in the
two point case, ie. $L_0^{(1)}=L_0^{(2)}$.

The next simplest representation contains a
singular vector at level two: $(L_{-1}^2-a L_{-2})\is{\vf _3}=0$
where $a={4h_3+2\over 3}$.
The action of the generator $L_0^{(1)}$ on the two dimensional
space spanned by $v=\is{\vf_1}\otimes\is{\vf_2}\otimes\is{\vf_3}$
and $L_{-1}^{(3)}v$  is given by
\bea
L_0^{(1)}v &=& z_{21}^{-1}\left \{ (z_{21}h_2+(z_{31}+z_{32})h_3)v+
z_{31}z_{32}L_{-1}^{(3)}v \right \} \nm
L_0^{(1)}L_{-1}^{(3)}v &=& z_{21}^{-1}\left \{ 2h_3 v+
 (z_{21}h_2+(z_{31}+z_{32}(h_3+1))L_{-1}^{(3)})v+
z_{31}z_{32}(L_{-1}^{(3)})^2v \right \}
\eea
Now we can replace $(L_{-1}^{(3)})^2v$ with $aL_{-2}^{(3)}v$ 
and use 
\be
L_{-2}^{(3)}=z_{32}^{-1}L_{-1}^{(2)}+z_{32}^{-2}L_{0}^{(2)}+ \dots
 +z_{31}^{-1}L_{-1}^{(1)}+z_{31}^{-2}L_{0}^{(1)}+ \dots ,
\ee
which is a consequence of the definition of the conformal blocks. 
All the relations are valid when they act on conformal blocks on the
right.  Using the global 
transformations we can express $L_{-1}^{(2)}$ and $L_{-1}^{(1)}$
in terms of the zero modes and $L_{-1}^{(3)}$ and $ L_{1}^{(3)}$,
the result is the following matrix:
\be
L_0^{(1)}=z_{21}^{-1} \left (
\ba{cc} 
\ba{l} z_{21}h_2+\\ (z_{31}+z_{32})h_3 \ea & 2h_3 +a 
         \left \{ \ba{l} h_2(z_{32}^{-1}z_{31}-2z_{21}^{-1}z_{31}
                      -2z_{12}^{-1}z_{32}+z_{32}z_{31}^{-1}) \\
                 h_3(-2z_{31}^2z_{21}^{-1}+2z_{31}z_{32}z_{12}^{-2}
                      +z_{32}z_{21}^{-1}+z_{31}^{-1}z_{21}^{-1}z_{32}^2)
           \ea \right \} \\
z_{31}z_{32} & \ba{l} z_{21}h_2+(z_{31}+z_{32})(h_3+1)+\\a 
                     (-z_{21}^{-2}z_{31}^3+z_{12}^{-2}z_{32}^2z_{31}+
                        2z_{21}^{-1}z_{32}^2) \ea
\ea \right ) 
\ee
This matrix looks a bit complicated, however it turns out that its 
eigenvalues are $z$-independent. This leads to take the $z_1 \to 
\infty$, $z_2 \to 1$ and $z_3\to 0$ limit. In this case the matrix 
takes the following simple form:
\be
L_0^{(1)}= \left (
\ba{cc}
h_2+h_3 & -ah_2 \\
-1 & h_2+h_3+1-a 
\ea  \right ) \nonumber 
\ee
Now if we use the standard parametrisation, $h(r,s)=((rt-s)^2-(t-1)^2)/4t$,
then $h_3=h(2,1)$ and for $h_2=h(r,s)$ the eigenvalues become 
$h(r+1,s)$ and $h(r-1,s)$. This is very similar to the fusion of the 
fundamental representation of $sl_2$.

Since the eigenvalues are $z$-independent we can take the limit above
at the beginning. The resulting Ward identities can be described as
\bea
L_0^{(2)}&=&L^{(3)}_{-2}-L^{(3)}_{-1} \nm
L_0^{(1)}&=&L^{(3)}_{-2}-2L^{(3)}_{-1}+L^{(3)}_{0} \\
0 &=& L^{(3)}_{-n}-2L^{(3)}_{-n+1}+L^{(3)}_{-n+2} \sp n>2 
\nonumber 
\eea
These relations hold when acting on conformal blocks on the right. 
These are the operators used by Feigin and Fuchs in \cite{FeFu}. 
They considered a given representation $\vf_3$ and analysed the 
possible values for $L_0^{(1)}$ and $L_0^{(2)}$. The restriction 
comes from the decoupling of the singular vectors of the third 
representation, which can be expressed as a polynomial in 
$L_0^{(1)}$ and $L_0^{(2)}$.

The $z$-dependence of the correlation functions can be described 
in the following way:
The configuration space where the arguments of the correlation 
function take values are $C_N=\left \{
\{z_1, \dots z_N \} \in \S ^N\s z_i\neq z_j \right \}$. Now take 
an open neighbourhood, $U$, of a point in the configuration space 
and associate with it a holomorphic dual to $V$ as 
\be
V^*(U)=\left \{ u_{.} :U\to V^*,\ {\rm holomorphic}: \ \ 
 {\rm that \  is} \quad  u_{z_1, \dots z_N}(v)\quad {\rm holomorphic}
  \quad \forall v\in V \right \} .
\ee
The meromorphic test-functions naturally extends to $U$, ie. they have
poles only when their argument coincides with one of the $z_i$-s, 
$\{z_1, \dots, z_N\} \in U$. This space is denoted by $\W^2(U)$.  
The space of the conformal blocks associated to the open set $U$ and the 
representation $V$ can be defined as those elements of the holomorphic dual 
which satisfy the Ward identities:
\be
E(U)=\left \{ u\in V^*(U) \vert \quad u(xv)=0 \sp \forall v\in V \ 
, \ x\in \W^2(U) \right \}  .
\ee 
The space of the conformal blocks can be regarded as a holomorphic vector 
bundle 
over the moduli space of the punctured sphere. 
The last thing we have to demand is that the operator  $L_{-1}$
to be the operator of the translation on the conformal blocks. 
This can be achieved by defining a flat connection on the space of the 
conformal blocks. We 
define the covariant differentiation,   $\nabla=
\sum_i\nabla_{z_i}\otimes dz_i$, as 
\be
\nabla_{z_i}u(v)=\p _{z_i} u(v)-u(L_{-1}^{(i)}v) .
\ee
for each $v\in V$. From the fact that $\nabla_{z_i}(ux)=(\nabla_{z_i}u)x+u\p _{z_i}x$
for $x \in \W^2(U) $ it follows, that $\nabla $ maps the space $E(U)$ into 
$\Omega^1(U)\otimes E(U)$, ie. it preserves the space of the conformal blocks. 
We note that the connection is flat: 
\be
[\nabla_{z_i},\nabla_{z_j}]=0.
\ee 
Now the conformal blocks are defined to be those sections of $E(U)$
which satisfy the horizontality condition: 
\be
\nabla u=0. 
\ee
These equations in the Kac-Moody case are nothing but the famous Knizhnik-
Zamolodchikov equations, if we represent the Virasoro algebra via the 
Sugawara construction.

\section{ The \wa2  theory}

Now let us try to apply the techincs introduced above for the 
simplest nontrivial $W$-theory, namely for a theory having the \wa2
algebra as the symmetry algebra. We define the space of the conformal
blocks in analogy with the Virasoro case. Since the symmetry algebra 
contains the Virasoro algebra as a subalgebra the previous 
considerations can be adapted for this case. The only new constraint we have to 
take into account comes from the Ward identities of the $W(z)$ current, 
(\ref{wardw}). 

More precisely, consider the Riemann sphere, $\S$, coordinatized by 
$z$ and $w=1/z$ and marked with $N$ fixed points, $\{z_i\}_{i=1}^N$. 
For each point we associate a h.w. representation of the \wa2 
algebra and take the tensor product 
$V=V_1\otimes \dots \otimes V_i \otimes \dots \otimes V_N$. 
We have the space of functions $\W^2$ as before, however now we define
an other space of functions $\W^3(\S \setminus \{p_1,\dots p_n \})$ --
they correspond to the $W$-transformations -- to be the space of complex
valued meromorphic functions, $f(z)$, for which $f(z) dz^{-2}$ is 
holomorphic except the points $p_1,\dots ,p_N$. (Clearly the 
globally defined $W$-transformations corresponds to the functions
$1,z,z^2,z^3,z^4$). 

Now we associate to each function, $f(z)$ of 
$\W^3(\S \setminus \{p_1,\dots p_n \})$ an action of the $W(z)$ 
field on $V$ in the following way. Make a Taylor expansion for 
$f(z)$ around the points $z_i$ and for the $(z-z_i)^{k+2}$ 
term associate the action of $W_k$ on $V_i$ thus the action 
of $W_n^{(i)}=1\otimes \dots \otimes W_n \otimes \dots \otimes 1$ 
on $V$. This map reads explicitly as:
\be
(z-z_i)^{-n}\rightarrow W^{(i)}_{-n-2}+ \sum_{l:l\neq
i}  \tilde W^{(l)}_{-n-2}  ,
\ee
where we have to make a difference depending on whether $n>0$ or $n\leq 0$:
\be
\tilde W^{(l)}_{-n-2}=\left \{ \ba {ll} (-1)^n \sum_{k=0}^\infty
z_{il}^{-n-k} 
{\scriptstyle{n+k-1\choose n-1}}W^{(l)}_{k-2} & n>0 \\
(-1)^n \sum_{k=0}^n z_{il}^{n-k} {\scriptstyle{n \choose
k}}W^{(l)}_{k-2} & n\leq 0  \ea \right .    .
\ee
We note that the generators with tilde all annihilate the vacuum. 

Similarly to the Virasoro case we define the space of conformal blocks 
associated to the points $\{ z_1, \dots, z_N\}$ and representation $V$ 
as the space of linear functionals on $V$ which is annihilated by $\W^2$ 
and $\W^3$, ($\W$ for short):
\be
E\left (V,\{p_i\}\right )=\left \{ u\in V^* \ \vert \ u(xv)=0 \sp 
\forall v\in V \ , \ x\in \W \right \} ,
\ee 

Before defining the $z$-dependence of the conformal blocks, we 
analyze the dimension of their space.  In order to 
describe the h.w. representation spaces, spanned by the ordered 
monoms of the generators with negative indices, we introduce some 
subalgebras of the \wa2 algebra. Define $ W_s$ to be the 
subalgebra spanned by those negative modes that annihilate the
vacuum, ie. it is generated by 
 $ W_s=\{ L_{-1},W_{-1},W_{-2} \}$. The other negative modes 
 generate another subalgebra which is denoted by $ W_{--}$. Now a 
 general element of the Verma modul $V_i$ can be written as
\be
v_i={ \cal W} _I\is{\vf_i}=
W^{j_1}_{-n_1}\dots W^{j_k}_{-n_k}W^{i_1}_{-m_1}\dots
W^{i_l}_{-m_l}\is{\vf_i} \sp W^{j_p}\in  W_{--} \s W^{i_p}\in 
 W_s 
\ee
where in the subalgebras defined above we choose the following 
ordering $j_l<j_{l+1}$ and if $ j_l=j_{l+1}$ then  $n_l\geq n_{l+1}$,
($i\lra j \s n\lra m$) and we introduced a compact notations for the 
genartors: $L=W^2\s W=W^3 $. The h.w. representation spaces, as 
in the Virasoro case are graded by the eigenvalue of $L_0$. 

Similarly to the Virasoro case it can be shown that the value of 
a conformal block on any vector is determined by its value on the vectors
generated by the generators of $ W_s$. The proof is analogous to 
the Virasoro case: if one of the representations contains a mode from 
$W_{--}$ then by the aid of the defining relation of the conformal blocks
we can express the value of any conformal block in terms of its other 
values on lower level vectors only. We describe this procedure in more
detail in the general $N$-point case. 

Focusing on the two point functions the value of the conformal blocks has to 
be defined on the vectors
\be
v=(L_{-1})^{i_1}(W_{-2})^{i_2}(W_{-1})^{i_3}\is{\vf_1}\otimes
(L_{-1})^{j_1}(W_{-2})^{j_2}(W_{-1})^{j_3}\is{\vf_2}
\ee
Besides the constraints coming form the global conformal transformation, 
(\ref{vir2}), we have the constraints of the global $W$-transformations:
\bea
u\left ((W^{(1)}_{-2}+\tilde
W^{(2)}_{-2})v\right )&=&u\left ( (W^{(1)}_{-2}+W^{(2)}_{-2})v\right )  =0\nm
 u\left ((W^{(1)}_{-1}+\tilde
W^{(2)}_{-1})v\right )&=&u\left ((W^{(1)}_{-1}+W^{(2)}_{-1}+z_{21}W^{(2)}_{-2})v\right )
=0\nm  u\left ((W^{(1)}_{0}+\tilde
W^{(2)}_{0})v\right )&=&u\left ((W^{(1)}_{0}+W^{(2)}_{0}+2z_{21}
W^{(2)}_{-1}+z_{21}^2W^{(2)}_{-2})v\right ) =0 \\
 u\left ((W^{(1)}_{1}+\tilde
W^{(2)}_{1})v\right )&=&u\left ((W^{(1)}_{1}+W^{(2)}_{1}+3z_{21}
W^{(2)}_{0}+3z_{21}^2W^{(2)}_{-1}+z_{21}^3W^{(2)}_{-2})v\right ) =0\nm
 u\left ((W^{(1)}_{2}+\tilde
W^{(2)}_{2})v\right )&=& 
 u\biggl (( W^{(1)}_{2}+W^{(2)}_{2}+4z_{21}
W^{(2)}_{1}+\nm
&& \hskip 3cm+6z_{21}^2W^{(2)}_{0}+4z_{21}^3W^{(2)}_{-1}
+z_{21}^4W^{(2)}_{-2})v\biggr ) =0 \nonumber  .
\eea
They correspond to the  element $(z-z_1)^n \s n=0\dots 4$ of $\W^3$. 
Clearly we have analogous but not independent equations for $(z-z_2)^n \s
n=0 \dots 4$. 
{}From the constraints (\ref{vir2}) the mode 
$L_{-1}$ can be eliminated. Similarly, using the equations above the 
modes $W_{-1}$ and $W_{-2}$ can be removed by the aid of the combinations
 $ W^{(1)}_{1}+\tilde W^{(2)}_{1}-z_{12}(W^{(1)}_{0} 
+\tilde W^{(2)}_{0})$, $W^{(1)}_{1}+\tilde
W^{(2)}_{1}-3/2z_{12}(W^{(1)}_{0}  
+\tilde W^{(2)}_{0})$ and $1\lra 2$, respectively.
Considering a linear combination of the last three equations
which does not contain any negative modes, we obtain
\bea
&u\left( \left (W^{(1)}_{2}+\tilde W^{(2)}_{2}-2z_{12}(W^{(1)}_{1}+\tilde
W^{(2)}_{1})+z_{12}^2(W^{(1)}_{0}+\tilde W^{(2)}_{0})\right )v\right )=\nm
&= u\left ((W^{(1)}_{0}+W^{(2)}_{0})v\right )=(w_1+w_2)u(v)=0  ,
\eea
where $W_0\is{\vf_i}=w_i\is{\vf_i}$. This shows that the dimension of 
the space of the conformal blocks is one if the Virasoro weights 
coincide and the $W$-weights are opposite and in all other cases it is  
zero. 
  
Now consider the $N$-point functions with insertion points
$z_1, \dots, z_N$. The conformal block $u$, by definition, assigns 
a complex number for each element of the form
\be
v=W^{j_1}_{-n_1}\dots W^{j_k}_{-n_k}
\is{\vf_1}\otimes \dots \otimes  W^{i_1}_{-m_1}\dots
W^{i_l}_{-m_l}\is{\vf_N}, 
\ee
where we used the ordering introduced earlier. 
Now we will show that if $v$ contain modes from $W_{--}$ than we 
can express the value of the conformal block with its values on vectors
generated by the modes of $W_s$. Without loss of generality we suppose that 
$v=(W^{j})^{(i)}_{-j-n}v^{'} \s n\geq 0$.
We use the definition of the conformal blocks, $u(xv)=0$, 
for $(z-z_i)^{-n-1}$ and obtain $u(v)=-\sum_{k:k\neq
i}(\tilde W^{j})^{(i)}_{-j-n}v^{'}=u(v^{''})$. Clearly 
the level of $v^{''}$ is smaller than the level of $v$.
Proceeding the same way we can eliminate all the modes of $W_{--}$
and replace them with the modes of $W_s$. Since during the procedure 
the level of the state in question is always decreasing and we are 
working with h.w. representation spaces  the procedure terminates. 

In order to reduce further the space of conformal blocks we use 
the global $W$-transformations:
\bea
u\Bigl ((W^{(i)}_{-2}+\sum_{j:j\neq i}\tilde
W^{(j)}_{-2})v\Bigr )&=&u\Bigl ((W^{(i)}_{-2}+\sum_{j:j\neq i} 
W^{(j)}_{-2})v\Bigr ) =0\nm
u \Bigl ((W^{(i)}_{-1}+\sum_{j:j\neq i}\tilde
W^{(j)}_{-1})v\Bigr )&=&u\Bigl ((W^{(i)}_{-1}+\sum_{j:j\neq
i}(W^{(j)}_{-1}+z_{ji}W^{(j)}_{-2}))v\Bigr ) 
=0\nm u \Bigl ((W^{(i)}_{0}+\sum_{j:j\neq i}\tilde
W^{(j)}_{0})v\Bigr )&=&u\Bigl ((W^{(i)}_{0}+\sum_{j:j\neq i}(W^{(j)}_{0}+2z_{ji}
W^{(j)}_{-1}+z_{ji}^2W^{(j)}_{-2}))v\Bigr ) =0 \nm
u \Bigl ((W^{(i)}_{1}+\sum_{j:j\neq i}\tilde
W^{(j)}_{1})v\Bigr )&=&u\Bigl ((W^{(i)}_{1}+\sum_{j:j\neq i}(W^{(j)}_{1}+3z_{ji}
W^{(j)}_{0}+\nm
&& \hskip 3cm +3z_{ji}^2W^{(j)}_{-1}+z_{ji}^3W^{(j)}_{-2}))v\Bigr ) =0\\
u \Bigl ((W^{(i)}_{2}+\sum_{j:j\neq i}\tilde
W^{(j)}_{2})v\Bigr )&=&u\Bigl ((W^{(i)}_{2}+\sum_{j:j\neq i}(W^{(j)}_{2}+4z_{ji}
W^{(j)}_{1}+\nm
&&\hskip 1cm+6z_{ji}^2W^{(j)}_{0}+4z_{ji}^3W^{(j)}_{-1}
+z_{ji}^4W^{(j)}_{-2}))v\Bigr ) =0\nonumber    .
\eea
We determine the value of the conformal block $u$ on the vector
\be
v=(L_{-1})^{i_1}(W_{-2})^{i_2}(W_{-1})^{i_3}\is{\vf_1}\otimes
\dots \otimes (L_{-1})^{j_1}(W_{-2})^{j_2}(W_{-1})^{j_3}\is{\vf_N}
\ee
Proceeding the same way as  we did in the case of the two point function, 
we take the difference of the ${3\over 2}z_{ji}$ multiple of the third 
equation and the fourth equation.  This combination makes it possible 
to eliminate the mode $W^{(j)}_{-2}$ at position $j$ without increasing
the level at position $i$. Similarly taking the difference of the $z_{ij}$
multiple of the third equation and the fourth equation the mode 
$W^{(j)}_{-1}$ can be eliminated. Using an analogous procedure the modes 
$W_{-1}$ and $W_{-2}$ can be eliminated at the position $i$ and $j$ 
simultaneously. Now taking the combination
\be
 W^{(i)}_{2} +\sum_{j:j\neq i}\tilde
W^{(j)}_{2}-2z_{ij}\Bigl (W^{(i)}_{1}+\sum_{j:j\neq i}\tilde 
W^{(j)}_{1}\Bigr )+z_{ij}^2\Bigl (W^{(i)}_{0}+\sum_{j:j\neq i}\tilde
W^{(j)}_{0}\Bigl )
\label{elimw2}
\ee
we can get rid of the mode $W^{(k)}_{-2}$. 

All in all this means that the value of the conformal blocks on any vector is 
determined if it is given on the vectors:
\bea
v=&W^{j_1}_{-n_1}\dots W^{j_k}_{-n_k}
\is{\vf_1}\otimes \dots \otimes\is{\vf_i}\otimes \dots \nm
&\dots \otimes
\is{\vf_j}\otimes \dots \otimes (W_{-1})^l\is{\vf_k}\otimes \dots \otimes
  W^{i_1}_{-m_1}\dots W^{i_l}_{-m_l}\is{\vf_N} 
\eea

Concretely in the case of the three point function the value of the
conformal block $u$ should be knonw on the vector
$\is{\vf_1}\otimes\is{\vf_2}\otimes \left ( W_{-1}\right ) ^l \is{\vf_3}$.
There is an important difference between the purely Virasoro symmetric 
theories and the theories having \wa2 symmetry. In the second case 
the space of the three point functions is infinite dimensional. (We note 
that the independence of the $\is{\vf_1}\otimes\is{\vf_2}\otimes \is{\vf_3}$
three point function from $\is{\vf_1}\otimes\is{\vf_2}\otimes 
W_{-1}\is{\vf_3}$ can be seen even in the simplest minimal model). 

In the three point case, as in the Virasoro theory, we can analyze the possible 
nonvanishing couplings.  We leave the operator $W_0^{(1)}$ undetermined and 
eliminate the negative modes in the first two representations. Contrary to the 
previous case we cannot use the equation which contains $W_0^{(1)}$, so we keep 
the equation (\ref{elimw2}) for defining $W_0^{(1)}$. 
This shows that in the third representation 
space everything, which is generated by the elements of $W_s$ remains 
undetermined in the general case. If however, we have singular vectors which 
makes this space finite then we have a finite dimensional matrix for 
$W_0^{(1)}$, whose eigenvalues give the nonvanishing couplings. 
This matrix has a simple form if we take the limit $z_1\to \infty$,
$z_2\to 1$ and $z_3\to 0$. In this case the constraints, which annihalate
the conformal blocks on the right, are:
\bea
W_0^{(2)}&=&-W^{(3)}_{-3}+2W^{(3)}_{-2}-W^{(3)}_{-1} \nm
W_0^{(1)}&=&-W^{(3)}_{-3}+3W^{(3)}_{-2}-3W^{(3)}_{-1}+W^{(3)}_{0} \\
0 &=&-W^{(3)}_{-n}+3W^{(3)}_{-n+1}-3W^{(3)}_{-n+2}+W^{(3)}_{-n+3} \sp n>3 
\nonumber \eea
They can be used to analyze the fusion a la Feigin and Fuchs. This was 
performed by Watts,  see \cite{W3fus}
for the details. 

We investigate the possible couplings, similarly as we did in the 
Virasoro case,  by starting  with the simplest cases. Consider the 
representation space where the factorspace $V_s:=W_s v/ W_{--}v$
is one dimensional. This is the vacuum representation with
two singular vectors at level one which can be chosen as $L_{-1}\is{\vf_3}=
W_{-1}\is{\vf_3}=0$. We also have the descendant singular vector $W_{-2}
\is{\vf_3}=0$ which shows that the special subspace is one dimensional. 
This implies that the analysis of the three point function reduces to 
the analysis of the two point functions as in the Virasoro case. 

In the next simplest case we have a singular vector at level one and 
another one at level two. This corresponds to the Toda model \cite{A_2}.
In order to describe this representation space we parametrize 
the central charge of the \wa2 algebra as 
\be
c=2-24(\beta -\bm)^2   .
\ee
Now in this completely degenerate representation space we have a h.w. singular
vector at level one
\be
\left (W_{-1}-(5/6\beta+1/2\bm)L_{-1}\right )\is{u}=0 . 
\ee
We also have a combination of the h.w. singular vector at level two and 
a descendats of the h.w. singular vector above, which has a very simple form
\be
\left (W_{-2}-\bm L_{-1}^2+2/3\beta L_{-2}\right )\is{u}=0 . 
\ee
Moreover we have the Toda equation of motion which is a descendant singular 
vector at level three:
\be
\left (W_{-3}-\beta ^{-3}L_{-1}^3+\bm L_{-1}L_{-2}+(\beta
/6-1/2\bm)L_{-3}\right )\is{u}=0 . 
\ee
This shows that the special subspace in question is three dimensional.

For each pair of the h.w. representations of the $sl_3$ algebra labelled by 
the highest weights $\L(n_1,n_2)=n_1\l_1+n_2\l_2 \s n\lra m$,  
(where $\l_1,\l_2$ denote the fundamental
weights), a h.w. representation of the \wa2 algebra can be associated via 
the following formulae:
\be
h={1\over 3}(x_1^2+x_1x_2+x_2^2)-(\beta-\bm)^2
 \qquad w={1\over 27}(x_1-x_2)(2x_1+x_2)(x_1+2x_2), 
\ee
where $x_i=(n_i+1)\beta -(m_i+1)\bm=:(n_i,m_i)$. 
The vacuum corresponds to the trivial representation, $x_1=(0,0), 
x_2=(0,0)$, and the Toda field to the fundamental representation 
$x_1=(1,0)$ and $x_2=(0,0)$. Note that the dimension of the 
special subspace is the same as the dimension of the $sl_3$ representation.
(This correspondence is much more straightforward at the classical level). 
Now if 
the parameters of the second representations are $(n_1,n_2)$ then the three 
possible eigenvalues of the operators $L_0^{(1)}$ and $W_0^{(1)}$ correspond 
to the parameters $(n_1+1,n_2-1)$, $(n_1-1,n_2)$ and $(n_1, n_2-1)$,
respectively, \cite{A_2}, (where the $m$ values are unchanged). 
This is very similar to the fusion rules of the 
fundamental representation of $sl_3$. 

The $z$-dependence of the conformal blocks are defined in the same way 
as in the Virasoro case, the only novelty is that the meromorphic testfunctions 
corresponding to the $W$-transformations have to extended to the neighbourhood
$U$. The horizontality condition gives differential equations for the 
conformal blocks, which can be solved in the simplest case, for details
see \cite{A_2}.

\section{ The case of general $W$-algebras}

We now follow the previous line and generalise the results obtained to 
other $W$-algebras. Let $\S$ denote the Riemann sphere 
with coordinates $z$ and $w=1/z$ and $V=V_1\otimes V_2\otimes
\dots  \otimes V_N$ the product of the highest weight 
representations, $V_i$, of the chiral algebra associated to the point $p_i$. 
The chiral algebra is supposed to be 
a $W$-algebra with integer spin fields only.  Introduce the space 
of complex valued functions on the Riemann sphere, $\W^j(\S
\setminus \{p_1,\dots p_n \})$, for whose elements, $f(z)$, 
$f(z)dz^{-j+1}$, are holomorphic except the points $p_1,\dots ,p_n$.
The globally defined transformations correspond
to the functions $1,z,\dots,z^{2j-2}$. 
For each functions of this type we associate an action of the 
spin $j$ generator of the chiral algebra on $V$ in the following
way: We Laurent expand $f(z)$ around $z_i$ and for the term 
$(z-z_i)^{k+j-1}$ we associate the action of $W^j_k$ on $V_i$ thus 
the action of $(W^j)_k^{(i)}=1\otimes
\dots \otimes W^j_n\otimes \dots \otimes 1$ on $V$. Since the 
functions of $\W^j$ are of the form $(z-z_i)^{n} \ , \ 
n< 2j-1$ we give the action explicitly for them:
\be
(z-z_i)^{-n}\to W^{(i)}_{-n-j+1}\oplus \sum_{l:l\neq i} 
\tilde W^{(l)}_{-n-j+1}  ,
\ee
or depending on whether $n$ is positive or negative 
\be
\tilde W^{(l)}_{-n-j+1}=\left \{ \ba {ll} (-1)^n
\sum_{k=0}^\infty z_{il}^{-n-k}
{\scriptstyle{n+k-1\choose n-1}}W^{(l)}_{k-j+1} & n>0 \\
 (-1)^n \sum_{k=0}^n z_{il}^{n-k}
{\scriptstyle{n \choose k}}W^{(l)}_{k-j+1} & n\leq 0 \ea \right . .
\ee
As in the previous cases the generators with tilde contain modes
only which annihilate the vacuum. 

Now having set the stage we define the space of the conformal blocks 
associated to the representation $V$ and the points $p_1, \dots, p_N$, 
as the space of linear functionals on $V$ which are annihilated by $\W$, 
the collection of the $\W^j$-s:
\be
E\left (V,\{p_i\}\right )=\bigl \{ u\in V^* \vert \quad
 u(xv)=0 \sp \forall v\in V \ , \ x\in \W \bigr \} .
\ee

As a first step we determine the dimension of the space of the 
conformal blocks. We introduce a basis in the h.w. representation, 
$V_i$ of the $W$-algebra. By definition $V_i=U( W_-)v_i$, where $W_-$ 
contains the negative modes of the generating fields.  
We devide the negative modes of the $W$-algebra into two groops depending on 
whether they annihilate the vacuum or not. 
\be
 W_s=\left \{ W^j_{-n} \vert \quad j>n>0 \right \} \sp
 W_{--}=\left \{ W^j_{-n} \vert \quad n\geq j \right \}  .
\ee
Clearly the vacuum  module is $U( W_{--})\v$. The elements of $W_s$ 
generate global transformations. The Verma modules can be described 
as  $V_i =U( W_{--})U( W_s)v_i$, where in the subalgebras the 
generators are ordered in such a way that the generators with smaller 
spins stay before the highers or in the case of the 
same spins the more negative preceeds the others. 
As before we will show that the value of the conformal block $u$ 
on every vector $v$ of $V$ is completely determined by its value on the 
vectors generated by the elements of $ W_s$. This means that 
we have to give the value of $u$ on the tensor products, $V_s$, of the vectors,
$(V_i)_s=U( W_s)v_i$. The proof is inductive in the level defined on 
the Verma module in the usual way: the level of a vectors in $V$ is simply
the sum of its levels in $V_i$. Clearly at the lowest grades we have the 
elements of $V_s$. Now suppose that we have a vector $v$ not in $V_s$, 
without loss of generaltily it has the following form:
\be
v=W^{j_1}_{n_1}\dots
W^{j_r}_{n_r}v_1\otimes \dots \otimes
W^{i_1}_{m_1}\dots
W^{i_p}_{m_p}v_i\otimes \dots \otimes W^{k_1}_{l_1}\dots
W^{k_s}_{l_s}v_N=(W^{i_1})^{(i)}_{m_1}v^{'}
\ee
where $m_1\leq -i_1$. No we use the definition of the conformal 
blocks $u(xv)=0$ for the element $x=(W^{i_1})^{(i)}_{m_1}+\sum_{j:j\neq i} 
(\tilde W^{i_1})^{(j)}_{m_1}$, which correspond to the function 
$(z-z_i)^{m_1+i_1-1}\in \W^{i_1} $ and  obtain 
\be
u(v)=u \Bigl (\sum_{j:j\neq i} (\tilde W^{i_1})^{(j)}_{m_1}v^{'}
\Bigr )= u(v^{''})
\ee 
Since the level of $v^{''}$ is smaller than the level of $v$ the 
statement is proved.

Now we turn to the two point functions. The constraints coming from
the global transformations with test functions 
$(z-z_1)^n \s n=0,1,\dots 2j-2$ for the spin $j$ generators are:
\bea
u\left ((W^{(1)}_{-j+1}+\tilde
W^{(2)}_{-j+1})v\right )&=&u\left ((W^{(1)}_{-j+1}+W^{(2)}_{-j+1})v\right )
=0\nm u\left ((W^{(1)}_{-j+2}+\tilde
W^{(2)}_{-j+2})v\right )&=&u\left ((W^{(1)}_{-j+2}+W^{(2)}_{-j+2}+z_{21}
W^{(2)}_{-j+1})v\right )=0\nm
u\Bigl ((W^{(1)}_{0}+\tilde
W^{(2)}_{0})v\Bigr )&=& u\Bigl ((W^{(1)}_{0}+\sum_{i=0}^{j-1}{\scriptstyle
 {j-1\choose i}}z_{21}^i W^{(2)}_{-i})v\Bigr ) =0 \\
 u\Bigl ((W^{(1)}_{j-2}+\tilde
W^{(2)}_{j-2})v\Bigr )&=& u\Bigl ((W^{(1)}_{j-2}+\sum_{i=0}^{2j-3}
{\scriptstyle {2j-3\choose i}}z_{21}^i W^{(2)}_{j-2-i})v\Bigr ) =0\nm
 u\Bigl ((W^{(1)}_{j-1}+\tilde
W^{(2)}_{j-1})v\Bigr )&=& u\Bigl ((W^{(1)}_{j-1}+\sum_{i=0}^{2j-2}
{\scriptstyle {2j-2 \choose i}}z_{21}^i W^{(2)}_{j-1-i})v\Bigr ) =0\nonumber  ,
\eea
Now consider the last $j$ equations. Descarding the last one for a while
the others contain the negative modes linearly independently due to the 
binomial nature of the coefficients. As a consequence we can find such a 
combinations of these equations which contain the negative mode
$W^{(2)}_{-i}$ only and nothing else. Applying this equations
recursively the mode $W^{(2)}_{-i}$ can be eliminated. Doing the same
procedure for the other generators we arrive at the h.w. vector. 
Clearly using the $1\lra 2$ trick,  only the h.w. vectors remain in both
representations spaces. Now using also the last equation we get:
\bea
u\Bigl (\bigl ( W^{(1)}_{j-1}&+&\tilde W^{(2)}_{j-1}+(j-1)z_{12}
(W^{(1)}_{j-2}+\tilde
W^{(2)}_{j-2})+\dots \nm&+&{\scriptstyle {j-1 \choose
i}}z_{12}^i(W^{(1)}_{j-1-i}+\tilde W^{(2)}_{j-1-i}) 
+\dots +z_{12}^{j-1}(W^{(1)}_{0}+\tilde W^{(2)}_{0})\bigr )v\Bigr )=\nm
&=&u\Bigl ((W^{(1)}_{0}+(-1)^{j-1}W^{(2)}_{0})v\Bigr )=
(w_1+(-1)^{j-1}w_2)u(v)=0  .
\eea
This shows that the dimension of the space of the conformal blocks is
one if the eigenvalue of the even generators are the same and the odd
generators are opposite and in every other cases it is zero. 

Now consider the $N$ point functions. Clearly it is enough to take into
account the value of the conformal blocks on the vectors generated by 
the elements of $W_s$. The constraints 
from the global transformations are:
\bea
u\Bigl ((W^{(i)}_{-j+1}+\sum_{l:l\neq i}^N\tilde
W^{(l)}_{-j+1})v\Bigl )&=& u\Bigl (\bigl (W^{(i)}_{-j+1}+\sum_{l:l\neq
i}^NW^{(l)}_{-j+1}\bigr )v\Bigr ) =0\nm  u\Bigl ((W^{(i)}_{-j+2}+\sum_{l:l\neq
i}^N\tilde W^{(l)}_{-j+2})v\Bigl )&=&u\Bigl ((W^{(i)}_{-j+2}+\sum_{l:l\neq
i}^N(W^{(l)}_{-j+2}+z_{li}W^{(l)}_{-j+1}))v\Bigl ) =0\nm
u\Bigl ((W^{(i)}_{0}+\sum_{l:l\neq i}^N\tilde
W^{(l)}_{0})v\Bigl )&=& u\Bigl ((W^{(i)}_{0}+\sum_{l:l\neq
i}^N\sum_{i=0}^{j-1}{\scriptstyle {j-1 \choose i}}z_{li}^i
W^{(l)}_{-i})v\Bigl ) =0 \\  u\Bigl ((W^{(i)}_{j-2}+\sum_{l:l\neq i}^N\tilde
W^{(l)}_{j-2})v\Bigl )&=&u\Bigl ((W^{(i)}_{j-2}+\sum_{l:l\neq
i}^N\sum_{i=0}^{2j-3}{\scriptstyle {2j-3 
\choose i}}z_{li}^i W^{(l)}_{j-2-i})v\Bigl ) =0\nm
 u\Bigl ((W^{(i)}_{j-1}+\sum_{l:l\neq i}^N\tilde
W^{(l)}_{j-1})v\Bigl )&=&u\Bigl ((W^{(i)}_{j-1}+\sum_{l:l\neq
i}^N\sum_{i=0}^{2j-2}{\scriptstyle {2j-2 
\choose i}}z_{li}^i W^{(l)}_{j-1-i})v\Bigl ) =0\nonumber  , 
\eea
where we did not write out explicitly the spin of the generator. 
Now taking the same linear combination of the equations which we saw at
the two point case we can eliminate the negative modes of $W^{(l)}_n$
without increasing the level at the position $i$ and $l$. Doing so 
reductively all negative modes at $l$ can be eliminated. Clearly 
we can do the same for $i$. Moreover with the aid of the last equation
the mode $W^{(k)}_{-j+1}$ of the spin $j$ generator at position $k$
can be reexpressed by others. This means that the correlation function
should be defined on the vectors of the form:
\be
v=W^{j_1}_{n_1}\dots W^{j_r}_{n_r}\otimes \dots \is{v_i}\otimes \dots 
\otimes
W^{i_1}_{m_1}\dots W^{i_p}_{m_p}\is{v_k}\otimes \dots
\otimes \is{v_j}\otimes W^{k_1}_{l_1}\dots \dots \otimes
W^{k_s}_{l_s}\is{v_N}
\ee
where $-i_s+1<m_s<0$. 
Concretely in the three point case this means that we have to
define the conformal bloks on the vectors:
\be
v=\is{v_1}\otimes \is{v_2}\otimes W^{k_1}_{l_1}\dots 
W^{k_s}_{l_s}\is{v_3} \sp -l_i<k_i
\ee
which means that the symmetry does not fix the three point functions 
completely as it is fixed in the Virasoro case. 

Turning to the completely degenerete representation we can express 
$W_0^{(1)}$ as an operator which contain negative modes only in 
the third representation, moreover only the modes of $W_s$. 
This shows that the number of possible couplings is bounded 
by the dimension of the factorspace $V_s=W_s\is{\vf_3}/W_{--}\is{\vf_3}$
which was called the special subspace by Nahm. He also obtained 
the same result considering the fusion of a field with finite dimensional 
subspace \cite{Na}.
 
One feasible
way to characterise the $W$-algebras can be achieved by analysing 
their special subspaces. The simplest representation space 
is always the vacuum representation
which contains the same number of singular vectors as the number of the 
generating fields of the $W$-algebra in question. These singular vectors
are independent so they can be chosen in the following form:
$W^j_{-1}v=0$ for all the generating primary fields. The primary 
nature of the fields yields that we have descendant singular 
vectors of the spin $j$ generator of the  form $Ad_{L_{-1}^k}W_{-1}^jv\sim   
W^j_{-k-1}v=0\s k=1,\dots ,j-2$. 
These show that the dimension of the special subspace 
is one and accordingly the analysis of the three point function reduces
to the analyses of the two point function. 

In the next simplest representation we have one singular vector at level 
two and all the others at level one. We believe that this representation
characterizes the $W$-algebra as fundamentaly as the vacuum representation. 
As the vacuum representation is the trivial representation in the quantum
case the next simplest should be the fundamental representation at the 
quantum level. This is where the special subspace have the smallest 
nontrivial dimension and consequently the number of fields that appear 
in the fusion product is the smallest. That is why we call this representation
the quantum fundamental representation, since it shares this property 
with the fundamental representation of Lie algebras. (The importance of 
this representation was also observed by Hornfeck \cite{Ho1}). 

This connection is explicit in the case of the \wg algebras: One 
possible way to quantize the Toda theories is to quantise their 
equation of motions, \cite{A_2,C_2}. In this language the special
subspace has the same dimension as it has in the classical case which
we analyse in the next section.

The $z$-dependence of the correlation functions can be described 
very similarly to the Virasoro case. Take a point in the configuration
space $C_N=\left \{ \{z_1, \dots z_N \} \in \S ^N\s z_i\neq z_j \right \}$
and for its neighbourhood, $U$, associate a holomorphic dual to $V$ as 
\be
V^*(U)=\left \{ u_{.} :U\to V^*,\ {\rm holomorphic}: \ \ 
 {\rm that \  is} \quad  u_{z_1, \dots z_N}(v)\quad {\rm holomorphic}
  \quad \forall v\in V \right \} .
\ee
Extend the meromorphic test-functions corresponding to the $W$-transformation
of the spin $j$ generator to $U$ and denote this space by $\W^j(U)$.  
The space of the conformal blocks associated to the open set $U$ and the 
representation $V$ are defined to be those elements of the 
holomorphic dual which satisfy the Ward identities:
\be
E(U)=\left \{ u\in V^*(U) \vert \quad u(xv)=0 \sp \forall v\in V \ 
, \ x\in \W(U) \right \}  .
\ee 
(Here $\W(U)$ denotes the collection of the $\W^j(U)$-s). 
Similarly to the Virasoro case the space of the conformal blocks can be 
regarded as a holomorphic vector bundle 
over the moduli space of the punctured sphere.
The $z$-dependence is defined by the aid of a flat connection: 
$\nabla=
\sum_i\nabla_{z_i}\otimes dz_i$, where 
\be
\nabla_{z_i}u(v)=\p _{z_i} u(v)-u(L_{-1}^{(i)}v) .
\ee
for each $v\in V$. Since $\nabla_{z_i}(ux)=(\nabla_{z_i}u)x+u\p _{z_i}x$
for $x \in \W(U) $ it maps the space $E(U)$ into 
$\Omega^1(U)\otimes E(U)$, ie. it preserves the space of the conformal blocks. 
Now the conformal blocks are defined to be those sections of $E(U)$
which satisfy the horizontality condition: 
\be
\nabla u=0. 
\ee

\section{Classical Toda theory and singular vectors}

The classical $W$-algebras corresponding to the principal $sl_2$ embedding
were defined in \cite{BaFeFo1,BaFeFo2}. They are nothing but 
the symmetry algebras of the Toda models, the theories that arises
as reductions of the Wess-Zumino-Witten-Models. Now we summarize the 
results we need. One imposes first class constraints on the space of the 
Kac-Moody currents. The $W$-algebra is the gauge invariant part of the 
constrained phase space, ie. it is invariant under the gauge transformations 
generated by the constraints. 
The defining relations of the $W$-algebra are given
by the Dirac brackets, and can be calculated very easily, which is one 
consequence of the reduction procedure. More concretely the 
$W$-transformations can be implemented by an appropiately chosen, 
KM transformation, $J_{imp}(W^{i})$:
\be
\delta J_{fix}(W^i(y))=\int dx \e _i(x) \{ W^{i}(x), J_{fix}(W^{i}(y))\}
=[J_{imp}(y),J_{fix}(W^{i}(y))]+\p J_{imp}(y).
\ee
The current, $J_{imp}(W^{i})$, also defines an action of the $W$-algebra 
on the group element of the WZNW model as $\delta g(x_+)=J_{imp}(x_+)
g(x_+)$, where this space is constrained as $\p_+g=J_{fix}g$. 
All information on a given classical $W$-algebra are contained in the 
matrices $J_{fix}$ and $J_{imp}$.

Concentrating on the Virasoro case we have for the fundamental 
representation the following results:
\be 
J_{fix}=\left ( \ba {cc} 0 & L(x) \\
                   1 & 0         \ea \right ) 
=t_-+L(x)t_+
\ee
for the gauge fixed current and 
\be
J_{imp}=\left ( \ba {cc} {1\over 2}\e_0(x) & \e_+(x) \\
                   \e_-(x) &-{1\over 2}\e_0(x)         \ea \right ) 
=\e_-(x)t_-+\e_0(x)\rho+\e_+(x)t_+
\ee
for the KM iplementation, where 
\be
\e_0(x)= \e_-(x)^{'} \sp \e_+(x)=L(x)\e_-(x)-{1\over 2}\e_-(x)^{''}.
\ee
Considering the classical representation space we have to solve the
equation $J_{fix}(x)g(x)=g(x)^{'}$. Concentrating on the columns we
have
\be
\left ( \ba {cc} 0 & L(x) \\
                   1 & 0         \ea \right ) 
\left ( \ba {c} u_1(x) \\
                u_0(x)         \ea \right )=
\left ( \ba {c} u_1(x)^{'} \\
                   u_0(x)^{'}    \ea \right ) 
\ee
In detail it reads as 
\be
u_1(x)=u_0(x)^{'} \sp u_0(x)^{''}-L(x)u_0(x)=0
\label{clsa1}
\ee
The action of the $W$-transformation on the representation space described 
above is defined as $\delta_{\e}g(x)=J_{imp}(x)g(x)$, 
ie. 
\be
\delta_{\e} u_0(x)=\e (x)u_0(x)^{'}-{1\over 2}\e (x)^{'}u_0(x) ,
\ee
This means, that classically $u_0$ is a hw. primary field with weight
$-{1\over 2}$ and the equation (\ref{clsa1}) can be considered as a classical
singular vector at level two. This has a very simimilar structure as the 
quantum singular vector of the quantum defining representation. 

In the case of the \wa2 algebra starting from the fundamental representation
of $sl_3$ the gauge  fixed current reads as
\be
J_{fix}=\left ( \ba {ccc}   0 & L(x)& W(x) \\
                      1 & 0 & L(x) \\
                      0 & 1 & 0          \ea \right )=\\ 
t_-+L(x)t_++W(x)l_{++}=t_-+W^2(x)t_++W^3(x)l_{++}  .
\ee
The current $J_{fix}$ can be obtained by imposing the constraint 
$\delta_{\e^L,\e^W}J_{fix}=[J_{imp},J_{fix}]+J_{imp}^{'}$ for the
element
\be
\e ^L(x)t_-+\e ^L_0(x)\rho+\e^L_+(x)t_++\e^W_{++}(x)l_{++}+
\e^W_{+}(x)l_{+}\\
+\e^W_{0}(x)l_{0}+\e^W_{-}(x)l_{-}+\e^W(x)l_{--} . 
\ee
The result is 
\bea
&\e^W_-={1\over 2} (\e^W)^{'} \s \e^L_0=-(\e^L)^{'} \s
\e^W_0={1\over 6}(\e^W)^{''}-{1\over 3}L\e^W \nm
&\e^L_+=W\e^W-(\e^L)^{''}+L\e^L \s
\e^W_+=-{1\over 6}(\e^W)^{'''}+{5\over 6}L(\e^w)^{'}+{1\over 3}
L^{'}\e^W \nm 
&\e^W_{++}=W\e^L+{1\over 6}(\e^W)^{''''}-{7\over 6}(L(\e^W)^{''}
-L^{'}(\e^W)^{'})-{1\over 3}L^{''}\e^W+{2\over 3}L^2\e^W ,
\eea
and
\bea
\delta L&=&\bigl[\e^LL^{'}+2(\e^L)^{'}L-2(\e^L)^{'''}\bigr]+
\bigl[2\e^WW^{'}+3(\e^W)^{'}W\bigr] \nm
\delta W&=&\bigl[\e^L W^{'}+3(\e^L)^{'}W_2\bigr] \nm
&&+\bigl[\e^W\bigl(-{1\over 6}L^{'''}+{2\over
3}L(L)^{'}\bigr)+(\e^W)^{'}\bigl(-{3\over 4}L^{''}+{2\over
3}L^2\bigr)\nm
&&-{5\over 4}(\e^W)^{''}L^{'}-{5\over
6}(\e^W)^{'''}L+{1\over 6}(\e^W)^{(V)}\bigr]  .
\eea
The representations are defined on the solution space of 
\be
J_{fix}=\left ( \ba {ccc}   0 & L(x)& W(x) \\
                      1 & 0 & L(x) \\
                      0 & 1 & 0          \ea \right )
\left ( \ba {c}   u_2 \\
                     u_1\\
                     u_0    \ea \right )
=\left ( \ba {c}   u_2^{'} \\
                     u_1^{'} \\
                      u_0^{'}    \ea \right )  
\ee
via $\delta g=J_{imp}g$. Explicitly we have
\be
u_1=u_0^{'} \sp u_2=u_0^{''}-Lu_0 \sp u_0^{'''}-2Lu_0^{'}+(L^{'}+W)u_0=0.  
\ee
The transformation of the last component reads as 
\bea
\delta u_0=&&\e^Lu_0^{'}-(\e^L)^{'}u_0 \nm
         &&+{1\over 6}(\e^W)^{''}u_0 
           +{1\over 2}(\e^W)^{'}u_0^{'} +\e^W\left (u_0^{''}-{2\over
3}Lu_0\right ). 
\eea
We can read off that it is a Virasoro and a $W$-primary field, moreover
it corresponds to a completely degenerate representation, since we have 
two independent classical singular vectors 
$\{W_{-1},u_0 \}={1\over 2}u_0^{'}$ and 
$\{W_{-2},u_0 \}=u_0^{''}-{2\over 3}Lu_0$. 
Clearly they have the same structure we had in the quantum case. 
We expect that this structure naturally generates for other represenations
of the classical algebra. (See the appendix for one example). In the 
general case the $\d _+g=J_{fix}g$ relation shows that the classical 
special subspace has the same dimension as the dimension of the 
representaion of the underlying algebra. 

\section{Conclusion}

We defined the space of the conformal blocks for a general $W$-algebra 
as the space of those linear functionals on the product of the representation
spaces that is annihilated by  certain meromorphic $W$-transformations. In
determining the dimension of this space we observed the importance of a 
special subspace of the h.w. representation spaces. This space is a 
factorspace spanned by those negative modes of the $W$-algebra that annihilate
the vacuum modulo those that do not. Analysing the three point functions of 
a field -- with finite dimensional special subspace -- and any fixed h.w. 
primary field, the number of the fields that can have nonzero three point 
functions is bounded by the dimension of the special subspace. 
This number is supposed to be finite for all the soluble models. We can 
charecterize the $W$-algebras by their special subspaces. For the simplest 
module, the vacuum module, the special subspace is one dimensional. We called
the next simplest representation space, a representation where the dimension 
of the special subspace takes the smallest nontrivial values, the quantum 
fundamental representation since in the case of the \wg algebras this is the 
quantum analogue of the fundamental representation of $G$. This is supported
by the dimension of the special subspace and by the fusion of this field, 
moreover can be corroborated by exploiting the relation between the classical 
and quantum Toda models. A further elaboration of this relationship and the 
analysis of the geometry of the conformal blocks is work in progress. 

\vspace{1cm}

\section{ Appendix}

The \wbc2 algebra, which is generated by the energy momentum tenzor 
$W_2$ and a spin four current $W_4$, is one of the simplest $W$-algebras
\cite{HaTa,KaWa2,BouC2}. The singular vectors 
in the "quantum defining representation" are very simple, see \cite{C_2}
for the details. We have a h.w. singular vector at level one
\be
W_{-1}\is{u} +\beta^{ 1} L_{-1}\is{u}=0
\ee
Combining the h.w. singular vector of level two with one of the 
descendants of the h.w. singular vector above we have a singular vector at
level two of the form:
\be
W_{-2}\is{u}+\beta^{ 2}L_{-2}\is{u}
+\beta^{ 1 1}L_{-1}^2\is{u}=0. 
\ee
They have a descendant at level three which has a particularly nice form
\be
W_{-3}\is{u}+\beta^{3}L_{-3}\is{u}
+\beta^{12}L_{-1}L_{-2}\is{u}
                        +\beta^{111}L_{-1}^3\is{u}=0. 
\ee
Among the descendants at level four we can find the quantum equation
of motion of the $C_2$ Toda model:
\bea
&& W_{-4}\is{u}
+\beta^{4}L_{-4}\is{u}+\beta^{22}L_{-2}^2\is{u}+ \nm
&&+\beta^{13}L_{-1}L_{-3}\is{u}+\beta^{112}L_{-1}^2L_{-2}\is{u}
+\beta^{1111}L_{-1}^4\is{u}=0  .
\eea
This shows that the dimension of the special subspace is four dimensional
in this case. 
If we had considered the $B_2$ Toda model instead of the $C_2$ then we 
would have obtained analogous singular vectors at the first three levels. The 
singular vector at level four in the form above is missing, however
we have a singular vector at level five in the form:
\bea
&\tilde\beta^{14}L_{-1}L_{-4}u-{2\over (\k+1)}\tilde\beta L_{-1}W_{-4}u
+\tilde\beta W_{-5}u+\tilde\beta^{122}L_{-1}L_{-2}^2u
+\tilde\beta^{113}L_{-1}^2L_{-3}u \nm
 &\qquad +\tilde\beta^{1112}L_{-1}^3L_{-2}u
+\tilde\beta^{11111}L_{-1}^5u+\tilde\beta^{23}L_{-2}L_{-3}u
+\tilde\beta^{5}L_{-5}u=0  .\hskip 2cm
\eea
It is in aggrement with the fact that the dimension of the fundamental
representation of $C_2$ is four and the $B_2$ is five. The fusion of 
these fields follow the selection rules of the four and five dimensional
representations of the underlying algebra. See \cite{Wbc_2null} for 
the details. 

It is interesting to compare this quantum representations 
 to the classical ones. In the
simplest case, which corresponds to the defining representation of the 
$C_2$ algebra, the form of the constrained current is 
\be
J_{fix}=\left ( \ba {cccc}   0 & 3L(x)& 0& W(x) \\
                      1 & 0 & 4L(x)&0 \\
                      0 & 1 & 0 &-3L(x) \\
                       0 & 0 & -1 &0      \ea \right )\\ .
\ee
The transformation of the Toda field and consequently the classical 
singular vectors are:
\bea
\delta u_0 =&& \e_1u_0^{'}-{3\over 2}\e_1^{'}u_0 + \nm
             \hskip -1cm &&+\e_2(-u_0^{'''}+{41\over 50}W_2u_0^{'}+
           {27\over 100}W_2^{'}u_0)+\e_2^{'}({1\over 2}u_0^{''}
           -{23\over 100}W_2 u_0)-{1\over 5}\e_2^{''}u_0^{'}
           +{1\over 20}\e_2^{'''}u_0 \nonumber . 
\eea
The analoge investigation in the $B_2$ case gives
\be
J_{fix}=\left ( \ba {ccccc}   0 & 4L(x)& 0& W(x)&0 \\
                      1 & 0 & 6L(x)&0&-W(x) \\
                      0 & 1 & 0 &-6L(x)&0 \\
                       0 & 0 & -1 &0&-4L(x) \\
                        0 & 0 & 0 &-1 &0         \ea \right )  ,
\ee
for the constrained current and 
\bea
\delta u_0 =&& \e_1u_0^{'}- 2\e_1^{'}u_0 + \nm
                     &&\e_2(-u_0^{'''}+{16\over 25}Lu_0^{'}+
           {7\over 25}L^{'}u_0)+\e_2^{'}(u_0^{''}
           -{18\over 25}L u_0)-{3\over 5}\e_2^{''}u_0^{'}
           +{1\over 5}\e_2^{'''}u_0  \nonumber .
\eea
for the transformation of the classical Toda field.

\end{document}